\documentclass[%
superscriptaddress,
reprint,
showpacs,
aps,
pra,
]{revtex4-1}
\usepackage{graphicx}
\usepackage{dcolumn}
\usepackage{bm}
\usepackage{braket}
\usepackage{amsthm}
\usepackage{epstopdf}
\usepackage{amssymb}
\usepackage{amsmath}
\usepackage{bbold}
\usepackage{hyperref}

\theoremstyle{definition}

\begin{document}
\title{Practical resources and measurements for lossy optical quantum metrology}
\author{Changhun Oh}%
\email{v55ohv@gmail.com}
\affiliation{Center for Macroscopic Quantum Control, Department of Physics and Astronomy, Seoul National University, Seoul 08826, Korea}

\author{Su-Yong Lee}%
\email{papercrane79@gmail.com}
\affiliation{School of Computational Sciences, Korea Institute for Advanced Study, Hoegi-ro 85, Dongdaemun-gu, Seoul 02455, Korea}

\author{Hyunchul Nha}
\affiliation{Department of Physics, Texas $A\&M$ University at Qatar, Education City, PO Box 23874, Doha, Qatar}
\affiliation{School of Computational Sciences, Korea Institute for Advanced Study, Hoegi-ro 85, Dongdaemun-gu, Seoul 02455, Korea}

\author{Hyunseok Jeong}%
\affiliation{Center for Macroscopic Quantum Control, Department of Physics and Astronomy, Seoul National University, Seoul 08826, Korea}
\date{\today}

\begin{abstract}
We study the sensitivity of phase estimation in a lossy Mach-Zehnder interferometer (MZI) using two general, and practical, resources generated by a laser and a nonlinear optical medium with passive optimal elements, which are readily available in the laboratory: One is a two-mode separable coherent and squeezed vacuum state at a beam splitter and the other is a two-mode squeezed vacuum state. 
In view of the ultimate precision given by quantum Fisher information, we show that the two-mode squeezed vacuum state can achieve a lower bound of estimation error than the coherent and squeezed vacuum state under a photon-loss channel. We further consider practical measurement schemes, homodyne detection and photon number resolving detection (PNRD), to characterize the accuracy of phase estimation in reality and find that the coherent and squeezed vacuum state largely achieves a lower bound than the two-mode squeezed vacuum in the lossy MZI while maintaining quantum enhancement over the shot-noise limit. 
By comparing homodyne detection and PNRD, we demonstrate that quadrature measurement with homodyne detection is more robust against photon loss than parity measurement with PNRD. We also show that double homodyne detection can provide a better tool for phase estimation than single homodyne detection against photon loss.
\end{abstract}


\maketitle

\section{Introduction}
Quantum metrology aims at achieving high precision in obtaining information about a physical system using quantum resources and measurements \cite{giovannetti2004}.
One prominent example is the detection of a gravitational wave using the Michelson interferometer, which measures a tiny variation of path length in the interferometer signifying the existence of a gravitational wave \cite{abbott2016,dowling2008}.
In an optical setting, it is also an important task to estimate an unknown phase-shift in one arm of the Mach-Zehnder interferometer (MZI).
Under the constraint of average input energy $\bar{n}$, the error $\Delta^2\phi$ of phase estimation using classical states of light is bounded by the shot-noise limit (SNL), $\Delta^2\phi\sim1/\bar{n}$. It can be enhanced up to the Heisenberg limit (HL), $\Delta^2\phi\sim1/\bar{n}^2$, by using quantum states of light
that rely on nonclassical resources like squeezing and quantum entanglement \cite{giovannetti2004,caves1981}.
In quantum phase estimation, one intends to minimize the variance of the estimator for a fixed value of phase below the SNL by employing quantum resources \cite{braunstein1994,paris2009}.
It is well known that the HL can be achieved by a NOON state in which all $N$ photons exist in either mode $a$ or $b$ \cite{dowling2008}. Within the HL, the phase sensitivity can be further enhanced by the entangled states generated with cat states \cite{joo2011,joo2012}, multi-headed cat states \cite{lee2015}, or the generalized NOON-type states \cite{zhang13, knott2016, lee2016}, etc.
The quantum enhancement can be indefinitely high, e.g., some of the NOON-type states provide an arbitrarily small quantum Cram\'{e}r-Rao bound (QCRB) even with a finite input energy \cite{lee2016}.

From a practical point of view, however, we must further investigate on the robustness of such quantum enhancement against inevitable noise in realistic situations.
Among numerous noisy models in optical interferometry \cite{escher2011,demkowicz2015}, one particularly important example is a photon-loss channel.
In a lossy MZI using NOON-type states, it turns out that the state with more quantum enhancement in an ideal situation becomes more fragile against noise \cite{lee2015,lee2016}.
In addition to the NOON states, there are other theoretical proposals to achieve robustness against photon-loss,
including a class of path-entangled Fock states \cite{huver2008}, entangled states generated by injecting twin-Fock states into a 50:50 beam splitter \cite{datta2011}, and a general two-mode pure states with definite photon number $N$ \cite{dorner2009,demkowicz2009}. 
For two-photon states, Kacprowicz \textit{et al}.\cite{kacprowicz2010} showed experimentally that the general two-mode entangled state with $N=2$ is more robust against photon-loss than the NOON state with $N=2$.
Although the proposed entangled states are more robust against photon-loss than NOON states, they are hard to prepare in practice due to the required 
high nonlinearity and controlled-operations with additional modes.

Here we are interested in practical input resources which can be readily prepared and used under photon loss. 
Specifically, we consider two input resources generated by a laser and a nonlinear optical medium. The first one is a two-mode separable coherent and squeezed vacuum (CSV) state that becomes entangled after the first beam splitter in the MZI \cite{caves1981,pezze2008, seshadreesan2011, jarzyna2012, lang2013, DBS2013, Gao2016, gard2017, LWYJS17} and the second one is a two-mode squeezed vacuum (TMSV) state \cite{steuernagel2004, olivares2009, anisimov2010, plick2010, zhang2013,huang2016}. We investigate these two classes of states to identify their merits for phase estimation employing 
two practical measurement schemes, i.e. parity measurement with photon number resolving detection (PNRD) and quadrature measurement with homodyne detection (HD). 
We compare their performance in a lossy MZI in terms of not only quantum Fisher information characterizing the ultimate precision but also the estimation errors directly obtained from measurement schemes.
We quantify the estimation error from measurement $\hat{O}$ by
\begin{align}\label{error}
\Delta^2\phi_{\hat{O}}=\frac{\langle\hat{O}^2\rangle-\langle\hat{O}\rangle^2}{|\partial\langle\hat{O}\rangle/\partial\phi|^2}.
\end{align}
Although both measurement setups provide us with the HL in a lossless MZI, we show that the quadrature measurement is more robust than the parity measurement in the lossy MZI.

This paper is organized as follows.  In Sec. \ref{sec:NMZI}, we introduce a model of lossy MZI with a phase shifter. In Sec. \ref{sec:PIS}, we investigate the phase sensitivity of two practical input resources, CSV and TMSV states, in terms of the ultimate precision using QFI. We consider two specific measurement setups, parity measurement with PNRD and quadrature measurement with singe (double) HD, and address phase-sensitivity under different measurement schemes. We summarize our results in Sec. \ref{sec:con}.

\section{Lossy Mach-Zehnder Interferometer}\label{sec:NMZI}
Let us first consider a lossless MZI which consists of two 50:50 beam splitters and a phase shifter. 
After combining input beams at the first beam splitter, 
an unknown phase is encoded into the beam by a phase shifter $\hat{U}_{\phi}=e^{-i\phi\hat{a}^{\dag}\hat{a}}$. Then after recombining the beams at the second beam splitter, measurements in the output modes are performed to obtain the phase information. Finally, the measurement data are processed to estimate the unknown phase. The quantum dynamics in the interferometer can be described by transformations of mode operators as follows. The first beam splitter changes two mode operators as $\hat{a}\rightarrow(\hat{a}+\hat{b})/\sqrt{2}$ and $\hat{b}\rightarrow(\hat{b}-\hat{a})/\sqrt{2}$, while the phase shifter gives $\hat{a}\rightarrow e^{-i\phi}\hat{a}$ and $\hat{b}\rightarrow\hat{b}$. The second beam splitter changes the mode operators as $\hat{a}\rightarrow(\hat{a}-\hat{b})/\sqrt{2}$ and $\hat{b}\rightarrow(\hat{a}+\hat{b})/\sqrt{2}$.

\begin{figure}[t]
\centerline{\scalebox{0.32}{\includegraphics[angle=-90]{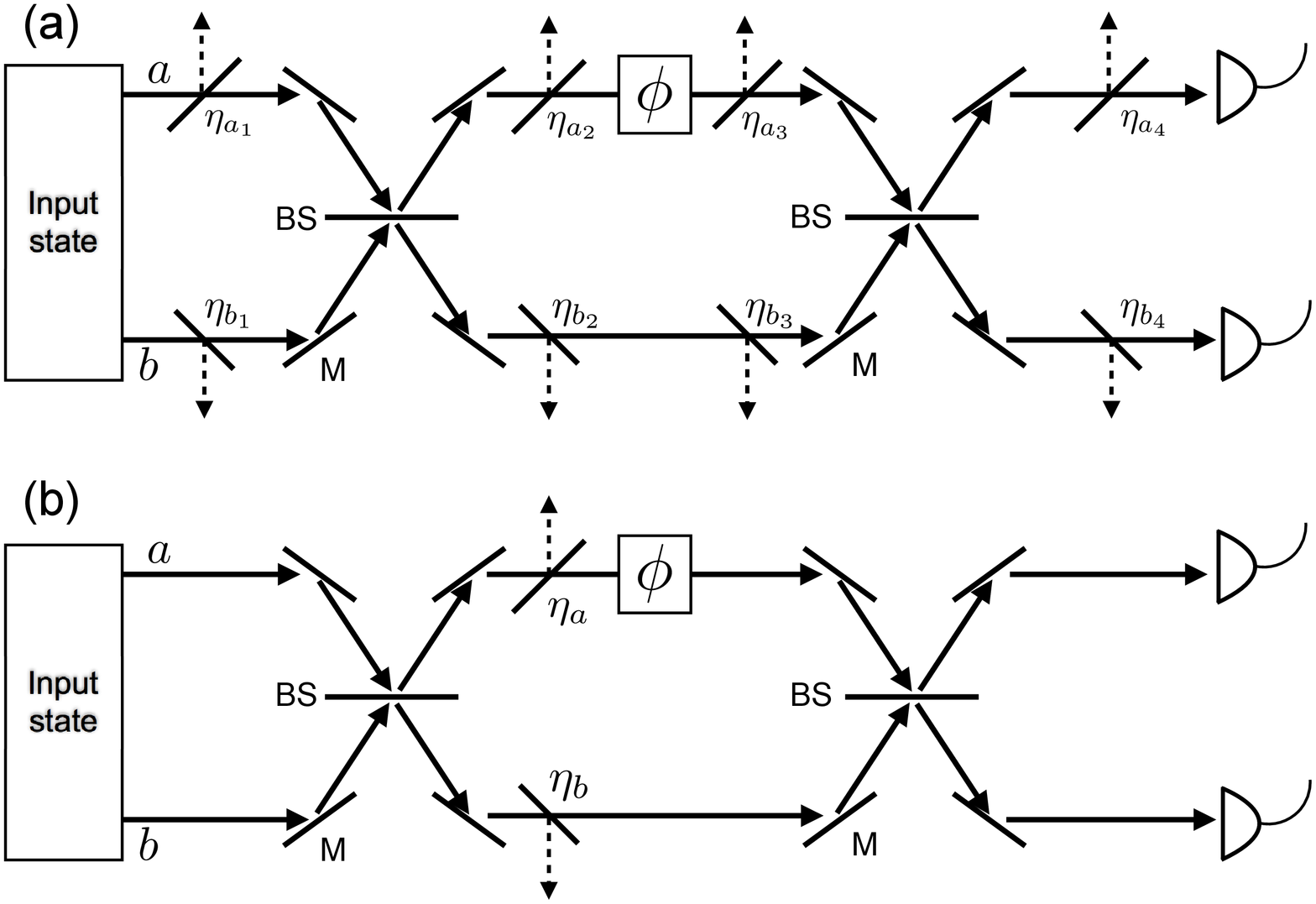}}}
\vspace{-0.3in}
\caption{Lossy Mach-Zehnder interferometer for phase estimation. (a) Photon loss in all possible paths and (b) simplified photon-loss model.  
Here $\eta$ is the transmittance of a fictitious beam splitter and
$\eta_a=\eta_p\eta_{a_2}\eta_{a_3}\eta_d$ and $\eta_b=\eta_p\eta_{b_2}\eta_{b_3}\eta_d$, where $\eta_{a_1}=\eta_{b_1}=\eta_p$ and $\eta_{a_4}=\eta_{b_4}=\eta_d$;
BS stands for a 50:50 beam splitter and $\phi$ a phase shift under the unitary action $\hat{U}_{\phi}=e^{-i\phi\hat{a}^{\dag}\hat{a}}$.
}\label{MZI}
\label{fig:MZI}
\end{figure}

We now consider photon loss by inserting fictitious beam splitters having a transmittance $\eta$ (a loss rate $1-\eta$) in an optimal path with a signal input beam and a vacuum state as two modes injected into the beam splitters \cite{leonhardt98}. In Fig.~\ref{MZI}(a), we place the fictitious beam splitters in all possible paths to consider photon-loss for all modes.
This configuration of a lossy interferometer can be simplified by noting that a phase shifting operation and a photon loss process commute \cite{demkowicz2009}. Thus the evolution loss characterized by transmissivities $\eta_{a_2} (\eta_{b_2})$ and $\eta_{a_3} (\eta_{b_3})$ can be combined as a single loss process. 
Moreover, if we assume the same preparation loss rate in each arm ($\eta_{a_1}=\eta_{b_1}=\eta_p$)
and the same detection loss rate in each arm ($\eta_{a_4}=\eta_{b_4}=\eta_d$), 
the preparation and detection losses can also be combined with the evolution loss so that we only need to consider two evolution losses with transmissivities $\eta_a=\eta_p\eta_{a_2}\eta_{a_3}\eta_d$ and $\eta_b=\eta_p\eta_{b_2}\eta_{b_3}\eta_d$. Note that we have also used the commutativity between a photon-loss process and a 50:50 beam-splitting process. 
Thus, with the above assumptions, all possible loss processes can be simplified into two evolution losses, as shown in Fig. \ref{fig:MZI}(b).

Although it is possible to consider various conditions on two loss rates, in this paper, we focus on two different situations for simplicity: (i) $\eta_a=\eta_b=\eta$ and (ii) $\eta_a=\eta$ and $\eta_b=1$. The former condition describes the case in which photon loss occurs symmetrically and the latter describes the case in which a noise occurs only along the paths of a phase shifter.

\section{Phase estimation with two practical input resources}\label{sec:PIS}
Let us consider two practical input resources for phase estimation. The first is a CSV state
\begin{align}
|\psi_{\text{CSV}}\rangle=\exp(\alpha \hat{a}^\dagger-\alpha^* \hat{a})\exp\bigg(\frac{1}{2}\xi \hat{b}^{\dagger2}-\frac{1}{2}\xi^*\hat{b}^2\bigg)|0\rangle,
\end{align}
where $\alpha=|\alpha|e^{i\theta_c}$ is the displacement parameter and $\xi=r e^{i\theta_r}$ the squeezing one.  The second practical input resource is a TMSV state
\begin{align}
|\psi_{\text{TMSV}}\rangle=\exp(\zeta \hat{a}^\dagger \hat{b}^\dagger-\zeta^*\hat{a} \hat{b})|0\rangle,
\end{align}
where $\zeta=s e^{i\theta_s}$ is the two-mode squeezing parameter. 
We note here that the phases of the considered states do not change optimal phase sensitivities but only shift the angles of optimal observables for homodyne detection, so we assume $\alpha$, $\xi$, and $\zeta$ to be real for simplicity.
The mean photon numbers of the states are given by $\bar{n}_{\text{CSV}}=\alpha^2+\sinh^2r$ and $\bar{n}_{\text{TMSV}}=2\sinh^2s$, which will be used as the energy constraint.
While the two states are known to achieve the HL without photon loss \cite{pezze2008, anisimov2010}, we investigate the phase sensitivity under practical situations with photon loss.
We particularly demonstrate our results with the mean photon number $\bar{n}=10$, since we obtain a similar tendency for different mean photon numbers (for example, $\bar{n}=7$ in Appendix A).

Note that CSV, TMSV, and coherent states that we investigate are Gaussian states \cite{ferraro2005, wang2007, marian2012, banchi2015} such that their characteristic function is given by
\begin{align}
\chi(\xi)=\text{Tr}[\hat{\rho} \exp(i\xi^T \hat{R})]=\exp\bigg[-\frac{1}{2}\xi^T\gamma\xi+id^T\xi\bigg],
\end{align}
where
\begin{align}
\gamma_{kl}&=\frac{1}{2}\langle\{\hat{R}_k,\hat{R}_l\}\rangle-\langle \hat{R}_k\rangle\langle \hat{R}_l\rangle, \label{cov} \\
d_k&=\text{Tr}(\hat{\rho} \hat{R}_k), \label{fm}
\end{align}
are the covariance matrix and the first-order moments, respectively. Note that $\xi\in \mathbb{R}^{4}$, $\hat{R}=(\hat{X}_a,\hat{P}_a,\hat{X}_b,\hat{P}_b)^T$, 
$\hat{X}_f=(\hat{f}+\hat{f}^\dagger)/\sqrt{2}$, and $\hat{P}_f=(\hat{f}-\hat{f}^\dagger)/\sqrt{2}i$ ($f=a,b$). 
 The formulas are widely used in our calculations.

\subsection{Quantum Fisher Information}
We first investigate the ultimate precision of the CSV and TMSV states by calculating quantum Fisher information. 
It is given by the QCRB as
\begin{align}
\Delta^2\phi\geq\frac{1}{M F_Q},
\end{align}
where $M$ is the number of trials repeated and $F_Q=\text{Tr}[\hat{\rho}_\phi \hat{L}_\phi^2]$ is the QFI of the state $\hat{\rho}_\phi$ containing phase information $\phi$. Here $\hat{L}_\phi$ is the so-called symmetric logarithmic derivative operator, which is given by the equation $\partial_\phi\hat{\rho}_\phi=(\hat{L}_\phi\hat{\rho}_\phi+\hat{\rho}_\phi \hat{L}_\phi)/2$ \cite{braunstein1994}. For a single-shot measurement ($M=1$), the inverse of QFI thus represents the lower bound for phase-estimation error. Using a phase shifting operation $\hat{U}_{\phi}=e^{-i\phi\hat{a}^{\dag}\hat{a}}$ in a lossless MZI,
we obtain the precision of a coherent state $\Delta^2\phi_{\text{SNL}}=1/2\bar{n}$ which sets the classical benchmark SNL.
Without photon-loss, we derive the QFIs for the CSV and TMSV states by using the covariance matrix and the first-order moment of the output mode (see Appendix B),
\begin{align}
F_Q^{\text{CSV}}&=\alpha^2e^{2r}+\sinh^2r+\alpha^2+\frac{\sinh^22r}{2} \nonumber \\ 
&\xrightarrow{\alpha=0}\bar{n}_{\text{CSV}}(2\bar{n}_{\text{CSV}}+3) \label{QFICSV}, \\
F_Q^{\text{TMSV}}&=8 \sinh^2s\cosh^2s=2\bar{n}_{\text{TMSV}}(\bar{n}_{\text{TMSV}}+2),
\end{align}
where $F_Q^{\text{CSV}}$ is maximized at $\alpha=0$ for a fixed mean photon number \cite{sparaciari2015}. The explicit expression of QFI of Gaussian states that we have used is provided in Appendix C. The QFIs show that both CSV and TMSV states attain the Heisenberg scaling. 
Note that for the other phase shifter $\hat{U}_{\phi}=e^{-i\phi(\hat{a}^{\dag}\hat{a}-\hat{b}^{\dag}\hat{b})/2}$
the QFIs take different forms \cite{pezze2008, lang2013, anisimov2010}, with the discrepancy discussed in Ref. \cite{jarzyna2012}.
In our work, we compare the ultimate bound from QFI and the achievable bounds by concrete measurement schemes under the phase shifter $\hat{U}_{\phi}=e^{-i\phi\hat{a}^{\dag}\hat{a}}$.

\begin{figure*}[t]
\centerline{\scalebox{0.62}{\includegraphics[]{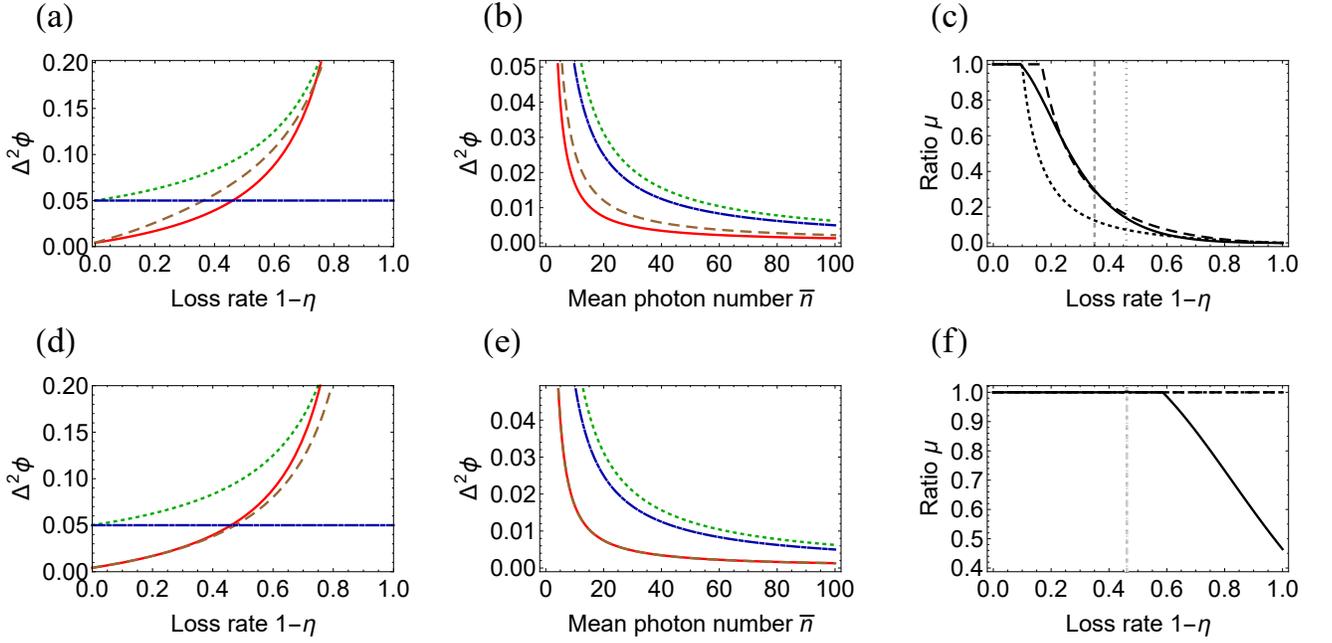}}}
\vspace{0.05in}
\caption{Quantum Cram\'{e}r-Rao bound for CSV (brown dashed curve), TMSV (red solid curve) and coherent (green dotted curve) states, respectively,  
as a function of loss rate $1-\eta$ at $\bar{n}=10$ (a) under symmetric photon-loss ($\eta_a=\eta_b=\eta$) and 
(d) under photon-loss in one arm ($\eta_a=\eta$ and $\eta_b=1$), and
as a function of $\bar{n}$ at a moderate loss rate $1-\eta=0.2$ (b) under symmetric photon-loss and (e) under photon-loss in one arm.
The shot-noise limit (blue dot-dashed lines) is given by $\Delta^2\phi_{\text{SNL}}=1/2\bar{n}$.
An optimal ratio $\mu$, which represents a portion of a single-mode squeezed vacuum state in the CSV state, is given
as a function of loss rate $1-\eta$ at $\bar{n}=1$ (black solid curve), $\bar{n}=10$ (black dashed curve), and $\bar{n}=100$ (black dotted curve) (c) under symmetric photon-loss and (f) under photon-loss in one arm. In (f), the dashed and dotted curves are overlapped.
In (c) and (f), the dashed (dotted) vertical line represents the loss rate to beat the SNL using the CSV state (TMSV state).
}
\label{fig:QFI}
\end{figure*}

In Fig. 2, we show the QCRB in the lossy MZI using the CSV and TMSV states.
The QFI in the lossy MZI is derived by following a method similar to the lossless case (see Appendixex B and C) and we provide the QFI of the CSV and TMSV states in Appendix C.
In the case of identical photon-loss in both arms ($\eta_a=\eta_b=\eta$), the TMSV state provides better phase sensitivity than the CSV state, as shown in Figs. 2(a) and 2(b). 
Specifically, for $\bar{n}=10$, the phase sensitivity of the TMSV state beats the SNL under the loss rate  $1-\eta< 0.46$ whereas the CSV state beats the SNL under $1-\eta<0.35$. 
In the case of photon loss only in one arm ($\eta_a=\eta$ and $\eta_b=1$), the phase sensitivities for both states are almost the same, as shown in Figs. 2(d) and 2(e). For $\bar{n}=10$, both states beat the SNL under the loss rate $1-\eta<0.46$.
We also compare the QCRBs of the CSV and TMSV states with that of coherent state under the same loss rate to verify if quantum enhancement still exists in the lossy interferometer. In Figs. 2(a) and 2(d), phase sensitivities of both states beat that of the coherent state unless the loss rate is too large. 
We thus achieve quantum enhancement using the CSV and TMSV states even in the lossy MZI.

For the case of CSV state, it is worth noting the fraction of the mean photon number of a single-mode squeezed vacuum state to the total mean photon number given by
\begin{align}
\mu\equiv\frac{\sinh^2{r}}{\alpha^2+\sinh^2{r}}.  \label{RATIO}
\end{align}
In a lossless MZI, we obtain the optimal ratio as $\mu=1$ from Eq.~\eqref{QFICSV}, i.e., injecting a single-mode squeezed vacuum state only is the optimal choice. 
In a lossy MZI, the optimal ratio becomes $\mu<1$ with the increment of loss rate, as shown in Figs. 2(c) and 2(f).
For a symmetric photon-loss, the optimal ratio decreases with the loss rate. On the other hand, for a photon loss in one arm, the optimal ratio approaches $\mu=1$ with the increment of the total mean photon number, regardless of the loss rate.

\subsection{Measurement setups}

In the preceding section we analyzed the ultimate theoretical estimation precision for the CSV and TMSV states in the lossy MZI by calculating QFI. We consider here specific measurement setups to examine the precision achievable in practice.

\subsubsection{Parity measurement with photon number resolving detection}
The first measurement setup to consider is the parity measurement with PNRD. The parity operator for the output mode $a$ is given by $\hat{\Pi}_a=(-1)^{\hat{a}^\dagger\hat{a}}$, which distinguishes between even and odd numbers of photons. The expectation value of the parity operator can be readily calculated by using the value of Wigner function at the origin, i.e. $\langle\hat{\Pi}_a\rangle=\pi W(0,0)$, where $W(x,p)$ is the Wigner function of the output mode $a$ \cite{barnett2002}. The Wigner function of a Gaussian state on the output mode $a$ is given by \cite{wang2007}
\begin{align}
W(x,p)=\frac{\exp[-\frac{1}{2}(X-d_a)^T \gamma_a^{-1} (X-d_a)]}{2\pi\sqrt{\det \gamma_a}},
\end{align}
where $X=(x,p)^T$. Here $\gamma_a$ and $d_a$ are the covariance matrix and the first-order moment of the state of the output mode $a$,

\begin{align}
\gamma_a=
\begin{pmatrix}
\gamma_{11} & \gamma_{12} \\
\gamma_{21} & \gamma_{22}
\end{pmatrix}
, \quad d_a=
\begin{pmatrix}
d_1 \\
d_2
\end{pmatrix}.
\end{align}
Thus the expectation value of the parity operator can be expressed as 
\begin{align}
\langle\hat{\Pi}_a\rangle=\frac{\exp(-\frac{1}{2}d_a^T \gamma_a^{-1} d_a)}{2\sqrt{\det \gamma_a}}.
\end{align}

\begin{figure}[t]
\centerline{\scalebox{0.38}{\includegraphics[]{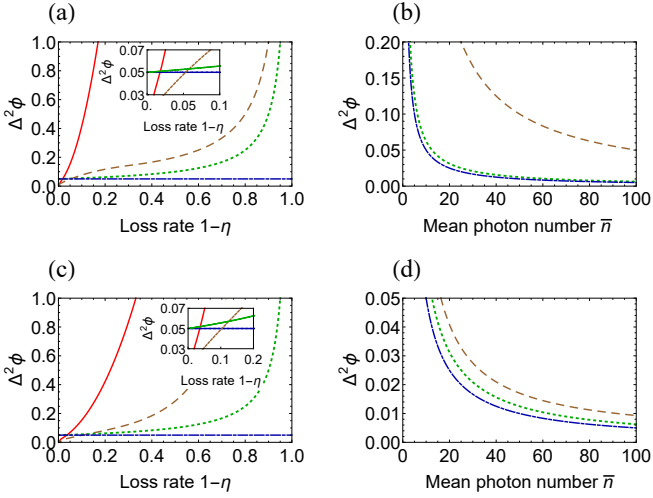}}}
\caption{
Phase sensitivity via parity measurement with photon number resolving detection, as a function of loss rate $1-\eta$ at $\bar{n}=10$ (a) under symmetric photon-loss ($\eta_a=\eta_b=\eta$) and 
(c) under photon-loss in one arm ($\eta_a=\eta$ and $\eta_b=1$) and
as a function of $\bar{n}$ at a loss rate $1-\eta=0.2$ (b) under symmetric photon-loss and (d) under photon-loss in one arm,
using CSV (brown dashed curve), TMSV (red solid curve) and coherent (green dotted curve) states.
The shot-noise limit (blue dot-dashed line and curve) is given by $\Delta^2\phi_{\text{SNL}}=1/2\bar{n}$.
 }
\label{fig:parity}
\end{figure}

Using the general expression for estimation error in Eq.~\eqref{error}, the phase sensitivity under parity measurement is given by
\begin{align}
\Delta^2\phi_{\hat{\Pi}_a}=\frac{\langle\hat{\Pi}_a^2\rangle-\langle\hat{\Pi}_a\rangle^2}{|\partial\langle\hat{\Pi}_a\rangle/\partial\phi|^2}=\frac{1-\langle\hat{\Pi}_a\rangle^2}{|\partial\langle\hat{\Pi}_a\rangle/\partial\phi|^2}.
\end{align}
In a lossless MZI, both the CSV and the TMSV states achieve the HL by using the parity measurement at an optimal angle $\phi$ \cite{seshadreesan2011, anisimov2010, seshadreesan2013},
\begin{align}
\Delta^2\phi_{\hat{\Pi}_a}^\text{CSV}&=\frac{1}{\alpha^2e^{2r}+\sinh^2{r}}\sim \frac{1}{\bar{n}_\text{CSV}^2},\\
\Delta^2\phi_{\hat{\Pi}_a}^\text{TMSV}&=\frac{1}{\bar{n}_{\text{TMSV}}(\bar{n}_{\text{TMSV}}+2)}.
\end{align}
In a lossy MZI, the phase sensitivity under parity measurement can also be calculated by inserting the photon-loss channel in the MZI and then deriving the covariance matrix and the first-order moment of the output state (see Appendix B).
While the parity measurement with PNRD attains the HL for the CSV and TMSV states under the lossless condition, the parity measurement can be extremely fragile against photon loss because single photon-loss distorts the parity information by changing the $(+$ or $-)$ sign in the parity operator. 
In Fig.~\ref{fig:parity}, we show that the results of the parity measurement are significantly degraded by photon loss. Furthermore, in contrast to the result on the QFI, the TMSV state is more fragile against photon loss than the CSV state under parity measurement. Although the CSV state is more robust than the TMSV state, it is also significantly fragile so that it becomes worse than the SNL even under a moderate loss rate $1-\eta>0.1$, thus the quantum advantage unexpected  for a small photon loss under parity measurement.

\subsubsection{Quadrature measurement with homodyne detection}
Although parity measurement performs well in a lossless phase estimation, we have shown that it is extremely fragile against photon loss. 
In this section, we consider another measuremenet, i.e., quadrature measurement with HD, to examine the sensitivity of phase-estimation in a lossy interferometer.
A balanced homodyne detection is used to measure the intensity difference between the two output modes generated by injecting a signal and a local oscillator field into a 50:50 beam splitter. The output data are used to obtain the expectation value of a field quadrature $\langle \hat{X}_{\varphi}\rangle=\Delta I /\sqrt{2}|\alpha_{\text{LO}}|$, 
where $\hat{X}_{\varphi}=\frac{1}{\sqrt{2}}(\hat{a}e^{-i\varphi}+\hat{a}^{\dag}e^{i\varphi})$, $\Delta I$ is the intensity difference, $\alpha_{\text{LO}}$ is the amplitude of the local oscillator field, and $\varphi$ is the phase of the local oscillator \cite{leonhardt98}.
From now on we represent two orthogonal quadratures as $\hat{X}_0=\hat{X}$ and $\hat{X}_{\pi/2}=\hat{P}$.

First we consider a quadrature measurement only in the output mode {\it a} by a single HD. 
In the lossless MZI, we find that the observables $\hat{P}$ for CSV states and $\hat{X}^2$ for TMSV states, respectively, provide the best phase sensitivity among all possible $\hat{X}_{\varphi}$ and 
$\hat{X}^2_{\varphi}$ observables.
For the CSV state, the first and the second moments of the field quadrature $\hat{P}_a$ are given by
\begin{align}
\langle\hat{P}_a\rangle^{\text{CSV}}&= d_2=\frac{\alpha\sin\phi}{\sqrt{2}},\\
\langle\hat{P}_a^2\rangle^{\text{CSV}}&= \gamma_{22}+d_2^2\\
&=\frac{1}{16}(4+4\cos\phi+3 e^{-2r}-4 e^{-2r}\cos\phi\nonumber\\
&+e^{-2r}\cos2\phi+2e^{2r}\sin^2\phi+8\alpha^2 \sin^2\phi). \nonumber
\end{align}

Here, we have used the characteristic function to calculate moments \cite{gardiner1991},
\begin{align}
\langle \hat{O}_{1} \hat{O}_{2} \cdot\cdot\cdot \hat{O}_{n}\rangle=\frac{1}{i^n}\frac{\partial^n}{\partial\xi_{\hat{O}_1}\partial{\xi_{\hat{O}_2}}\cdot\cdot\cdot\partial\xi_{\hat{O}_n}}\chi(\xi)|_{\xi=0},
\end{align}
where $\hat{O}_i \in \{\hat{X}_a,\hat{P}_a,\hat{X}_b,\hat{P}_b\}$ and $\chi(\xi)$ is the characteristic function of the output state.
Using Eq.~\eqref{error}, we obtain the phase sensitivity optimized over the angle $\phi$ as
\begin{align}
\Delta^2\phi_{\hat{P}_a}^{\text{CSV}}&=\frac{1}{\alpha^2 e^{2r}}\sim \frac{1}{\bar{n}^2_{\text{CSV}}}.
\end{align}
For the TMSV state, on the other hand, the first and the second moments of the field quadrature $\hat{X}_a$ are given by
\begin{align}
\langle\hat{X}_a\rangle^{\text{TMSV}}&=d_1=0, \\
\langle\hat{X}_a^2\rangle^{\text{TMSV}}&=\gamma_{11}+d_1^2\\
&=\frac{1}{2}(\cosh^2s+\sinh^2s-\sin^2\phi\sinh{2s}).\nonumber
\end{align}
In contrast to the CSV state, the first-order moment of the field quadrature $\hat{X}_a$ does not contain any phase information, so we choose the observable $\hat{X}_a^2$ as our signal of interest.
Using the higher-moment relations of Gaussian states \cite{gardiner1991}, we obtain the phase sensitivity at an optimal angle $\phi$ as
\begin{align}
\Delta^2\phi_{\hat{X}_a^2}^{\text{TMSV}}&=\frac{1}{2 e^{2s}}\bigg(\frac{1}{\sinh^2s}+\frac{1}{\cosh^2s}\bigg)\sim\frac{1}{\bar{n}_{\text{TMSV}}^2}.
\end{align}

\begin{figure}[t]
\centerline{\scalebox{0.38}{\includegraphics[]{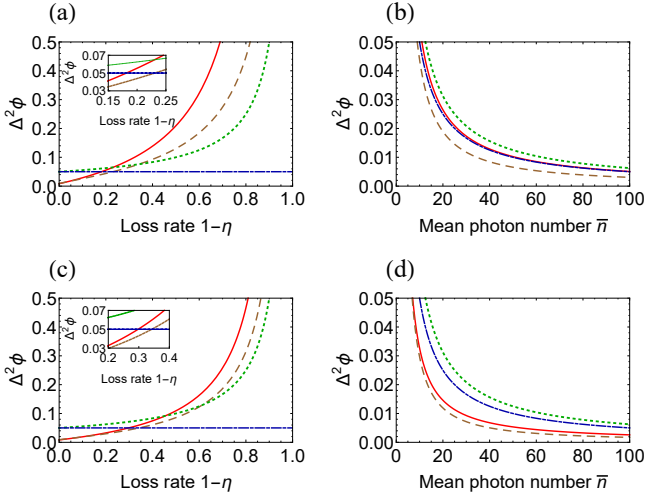}}}
\caption{
Phase sensitivity via quadrature measurement with single homodyne detection as a function of loss rate $1-\eta$ at $\bar{n}=10$ (a) under symmetric photon loss ($\eta_a=\eta_b=\eta$) and 
(c) under photon loss in one arm ($\eta_a=\eta,~\eta_b=1$), and
as a function of $\bar{n}$ at a loss rate $1-\eta=0.2$ (b) under symmetric photon loss and (d) under photon loss in one arm,
using CSV (brown dashed curve), TMSV (red solid curve) and coherent (green dotted curve) states.
The shot-noise limit (blue dot-dashed line and curve) is given by $\Delta^2\phi_{\text{SNL}}=1/(2\bar{n})$.}
\label{fig:shomodyne}
\end{figure}

Although both states achieve the Heisenberg scaling of phase sensitivity in an ideal situation, the quadrature measurement in one output mode only does not provide a better precision than parity detection under the lossless condition. In contrast, for a lossy MZI, the quadrature measurement with single HD provides a more robust phase sensitivity than parity measurement with PNRD, as shown in Fig. \ref{fig:shomodyne}.
Note that, in the lossy MZI, the phase sensitivity under quadrature measurement can be calculated by replacing the characteristic function of the output state in the lossless MZI with that in the lossy MZI (see Appendix B).
For the case of identical photon loss in both arms,  the CSV and the TMSV states attain better sensitivity than the SNL under the loss rate $1-\eta<0.23$ and $1-\eta<0.18$, respectively. For the case of photon loss only in one arm,  the CSV and the TMSV states beat the SNL under the loss rate $1-\eta<0.34$ and $1-\eta<0.3$, respectively. Furthermore, phase sensitivities of the CSV and TMSV states with single quadrature measurement are better than the QCRB of coherent state with the same condition unless the loss rate is too large. Thus, quadrature measurement with single HD enables us to achieve quantum enhancement in a lossy interferometer.

\begin{figure}[b]
\centerline{\scalebox{0.38}{\includegraphics[]{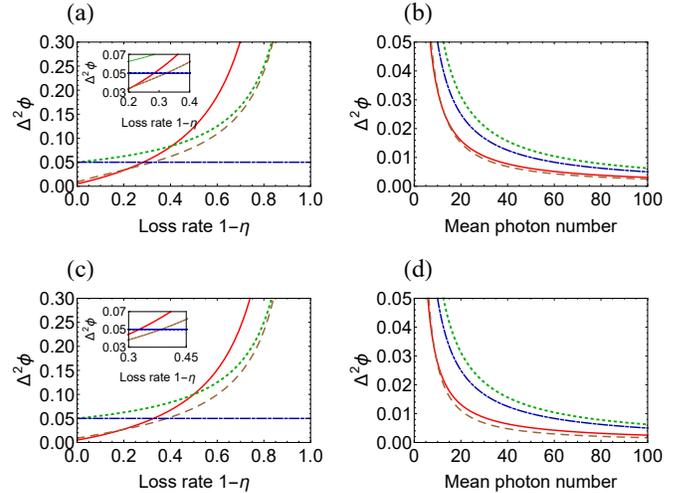}}}
\caption{
Phase sensitivity via quadrature measurement with double homodyne detection, as a function of loss rate $1-\eta$ at $\bar{n}=10$ (a) under symmetric photon loss ($\eta_a=\eta_b=\eta$) and 
(c) under photon loss in one arm ($\eta_a=\eta$ and $\eta_b=1$), and
as a function of $\bar{n}$ at a loss rate $1-\eta=0.2$ (b) under symmetric photon loss and (d) under photon loss in one arm,
using CSV (brown dashed curve), TMSV (red solid curve) and coherent (green dotted curve) states.
The shot-noise limit (blue dot-dashed line and curve) is given by $\Delta^2\phi_{\text{SNL}}=1/2\bar{n}$.
}
\label{fig:homodyne}
\end{figure}

We now consider quadrature measurements in both output modes by double homodyne detections for further enhancement of phase sensitivity. Under the lossless condition, phase sensitivities of CSV and TMSV states can be improved by using double HDs. 
It was previously proposed to use double HD for the TMSV state to achieve the Heisenberg scaling \cite{steuernagel2004}. Specifically, the scheme is a quadrature measurement of $\hat{X}$ in the output mode {\it a} and that of $\hat{P}$ in the output mode {\it b}, i.e., $\hat{O}=\hat{X}_a \hat{P}_b$. Here we find that for the TMSV state, $\hat{O}=\hat{X}_a\hat{X}_b$  provides better the phase sensitivity than $\hat{O}=\hat{X}_a \hat{P}_b$, and it is the optimal observable among all possible products of two quadratures.
In a lossless MZI with the TMSV state,  the first and the second moments of the field quadrature $\hat{X}_a \hat{X}_b$ are given by \cite{steuernagel2004}, 
\begin{align}
\langle\hat{X}_a\hat{X}_b\rangle^{\text{TMSV}}&=\gamma_{13}+d_1 d_3=\frac{1}{2}\sinh{2s}\cos^2\phi ,\\
\langle\hat{X}_a^2\hat{X}_b^2\rangle^{\text{TMSV}}&=4d_1 d_3 \gamma_{13}+2 \gamma_{13}^2+(d_1^2+\gamma_{11})(d_3^2+\gamma_{33})\nonumber \\
&=\frac{1}{64}\bigg[(17+4\cos2\phi)\cosh4s-1-4\cos2\phi \nonumber \\&+6\cos4\phi\sinh^22s-16\sin^2\phi\sinh4s\bigg].
\end{align}
Using Eq.~\eqref{error}, we obtain the phase sensitivity optimized over the angle $\phi$ as
\begin{align}
\Delta^2{\phi}^\text{TMSV}_{\hat{X}_a\hat{X}_b}&=\frac{1}{\sqrt{2+2 e^{8s}}-e^{4s}-1}
\xrightarrow{s\gg1}\frac{(\sqrt{2}+1)}{4\bar{n}^2_{\text{TMSV}}}.
\end{align}
Thus the quadrature measurement with double HD provides the Heisenberg scaling for the TMSV state, which shows better performance than parity measurement with PNRD and quadrature measurement with single HD in the lossless MZI. 
On the other hand, the CSV state can also achieve the Heisenberg scaling with a different observable, e.g. the sum of two quadratures measured in each output mode
\begin{align}
\Delta^2{\phi}^\text{CSV}_{\hat{X}_{\varphi_a}+\hat{X}_{\varphi_b}}&=\frac{1}{\alpha^2(e^{2r}+1)}\sim \frac{1}{\bar{n}^2_{\text{CSV}}}.
\end{align}
where we have chosen optimal quadratures in each output mode.

In Fig.~\ref{fig:homodyne}, we show the phase sensitivity in the lossy MZI using double HDs. 
For the case of identical photon-loss in both arms,   
the CSV state beats the SNL under the loss rate $1-\eta<0.32$ and the TMSV state under the loss rate $1-\eta<0.28$. 
For the case of photon-loss in one arm,
the CSV beats the SNL under the loss rate $1-\eta<0.39$, and TMSV states beat the SNL under the loss rate $1-\eta<0.33$. 
For the CSV and the TMSV states, the double HD provides better robustness than the single HD.

\begin{figure}[t!]
\centerline{\scalebox{0.38}{\includegraphics[]{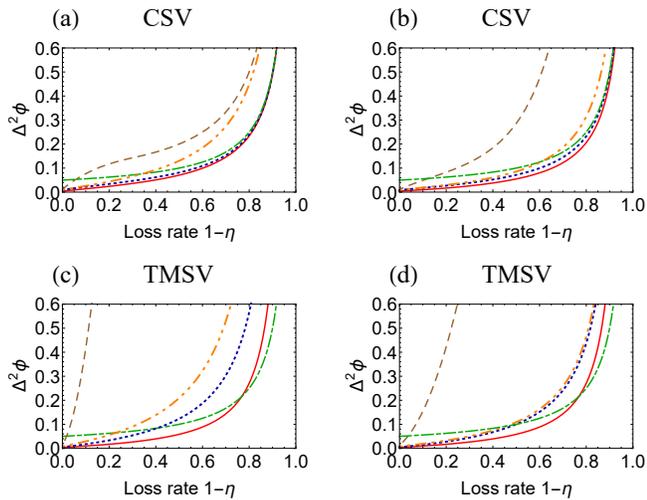}}}
\caption{
Comparison of phase sensitivity via the quantum Cram\'{e}r-Rao bound (red solid curve), parity measurement with photon number resolving detection (brown dashed curve), quadrature measurement with single homodyne detection (orange double-dot dashed curve) and double homodyne detection (blue dotted curve) at $\bar{n}=10$, using a CSV state (a) under symmetric photon loss and (b) under photon loss on one arm and a TMSV state (c) under symmetric photon loss and (d) under photon loss on one arm.  
The green dot-dashed curve represents the quantum Cram\'{e}r-Rao bound for a coherent state.}
\label{fig:comparison}
\end{figure}

To confirm quantum enhancement under a photon-loss channel, we compare the attainable precision limit under each measurement setup with the ultimate bound of the coherent state. 
In Fig.~\ref{fig:comparison}, we compare the phase sensitivities obtained by QFI, parity measurement with PNRD, and quadrature measurement with single (double) HD. 
We see that the quadrature measurement with HD provides more robust phase sensitivity than the parity measurement for both states, although it does not saturate to the ultimate QFI. The parity measurement with PNRD is extremely fragile in the lossy MZI.
In Figs. \ref{fig:comparison}(a) and \ref{fig:comparison}(b), the CSV state provides a quantum advantage using the quadrature measurement unless the loss rate is extremely high both under the symmetric photon-loss and under the photon-loss in one arm. 
In Figs. \ref{fig:comparison}(c) and \ref{fig:comparison}(d), the TMSV state also provides quantum advantage using the quadrature measurement, for the loss rate 
$1-\eta<0.4$ under the symmetric photon-loss and for the loss rate $1-\eta<0.47$ under the photon-loss in one arm, respectively. 
The CSV state maintains quantum enhancement better than the TMSV state both under the symmetric photon-loss and under the photon-loss in one arm.
In particular, phase sensitivity via double HD is better than that via single HD for both the CSV and the TMSV states under the photon-loss models.

\section{Conclusion} \label{sec:con}
In this work, we have investigated two practical input resources, coherent and squeezed vacuum state and two-mode squeezed vacuum state that are available in laboratory, for lossy optical quantum metrology. 
To characterize their usefulness for phase estimation, we considered both the quantum Fisher information giving the ultimate precision of phase estimation and the estimation errors directly obtained from practical measurement schemes, homodyne detection and photon number resolving detection.
We have found that the two-mode squeezed vacuum state provides a better resource in view of the ultimate precision given by quantum Fisher information than the coherent and squeezed vacuum state under (i) a symmetric photon loss ($\eta_a=\eta_b=\eta$). For the coherent and squeezed vacuum state, the optimized portion of a squeezed vacuum state against total energy of the state decreases with the photon-loss rate. Under (ii) a photon loss in one arm only ($\eta_a=\eta$ and $\eta_b=1$), we have obtained that the coherent and squeezed vacuum state can demonstrate better performance than the two-mode squeezed vacuum state. In this case, 
the optimized portion of a squeezed vacuum state becomes $1$ regardless of the loss rate.

On the other hand, under practical measurement setups considered (homodyne detection and PNRD), it has been shown that the coherent and squeezed vacuum state is more robust against photon loss than the two-mode squeezed vacuum state while maintaining quantum enhancement over the shot-noise limit. Comparing the parity and the quadrature measurements, we have shown that the quadrature measurement is more robust than the parity measurement and that the double homodyne detection exhibits better robustness than the single homodyne detection under the photon-loss channel.

In this paper, we have fixed the total mean photon number as $\bar{n}=10$. Under the current technology, it is possible to generate a two-mode squeezed vacuum state with $\bar{n}=10$.
In experiment, the generation of $15$-dB single-mode squeezed vacuum states was reported \cite{vahlbruch2016}, which corresponds to 
$\bar{n}\approx7$. Injecting each single-mode squeezed vacuum state with $\bar{n}=7$ into a 50:50 beam splitter, we can obtain the two-mode squeezed vacuum state with $\bar{n}=14$.
Although the coherent and squeezed vacuum state may not approach the range of $\bar{n}=10$,  we obtain phenomena similar to the results shown in this paper for the case of $\bar{n}=7$ (see Appendix A).

We have considered here the quadrature observables based on the first and the second moments. As a future work, it would be interesting to extensively consider higher-order moments of quadrature observable to enhance the phase sensitivity up to the quantum Cram\'{e}r-Rao bound in the lossy MZI. Moreover, we may incorporate the adaptive phase control method to achieve better performance under a practical measurement setting \cite{BW00}.

\section*{acknowledgments}
SYL would like to thank Changhyoup Lee for useful comments.
This work was supported by a National Research Foundation of Korea grant funded by the Korea government (MSIP) (No.\ 2010-0018295) and by the KIST Institutional Program (Project No.\ 2E26680-16-P025).
HN was supported by NPRP Grant No. 8-352-1-074 from Qatar National Research Fund.

\section*{Appendix}

\subsection{Phase sensitivities with $\bar{n}=7$}
\setcounter{equation}{0}
\setcounter{figure}{0}
\renewcommand{\theequation}{A\arabic{equation}}
\renewcommand{\thefigure}{A\arabic{figure}}

\begin{figure}[h]
\centerline{\scalebox{0.38}{\includegraphics[]{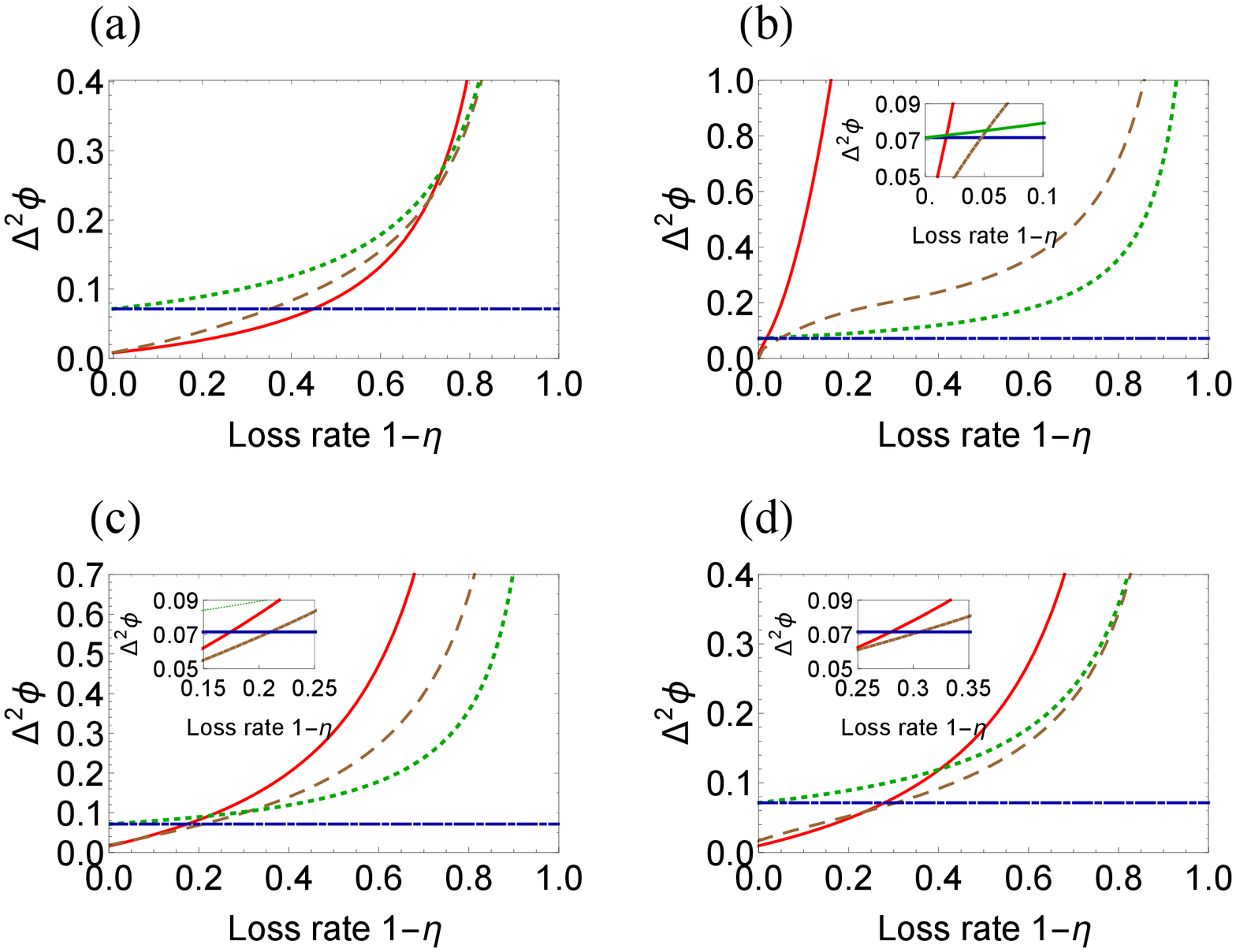}}}
\caption{Phase sensitivities for CSV (brown dashed curve), TMSV (red solid curve), and coherent (green dotted curve) states, respectively, as a function of loss rate $1-\eta$ 
at $\bar{n}=7$ under symmetric photon-loss via (a) the quantum Cram\'{e}r-Rao bound, (b) parity measurement with PNRD, (c) quadrature measurement with single homodyne detection, and (d) quadrature measurement with double homodyne detection.
The shot-noise limit (blue dot-dashed line) is given by $\Delta^2\phi_{\text{SNL}}=1/2\bar{n}$.
}
\end{figure}

\begin{figure}[h]
\centerline{\scalebox{0.38}{\includegraphics[]{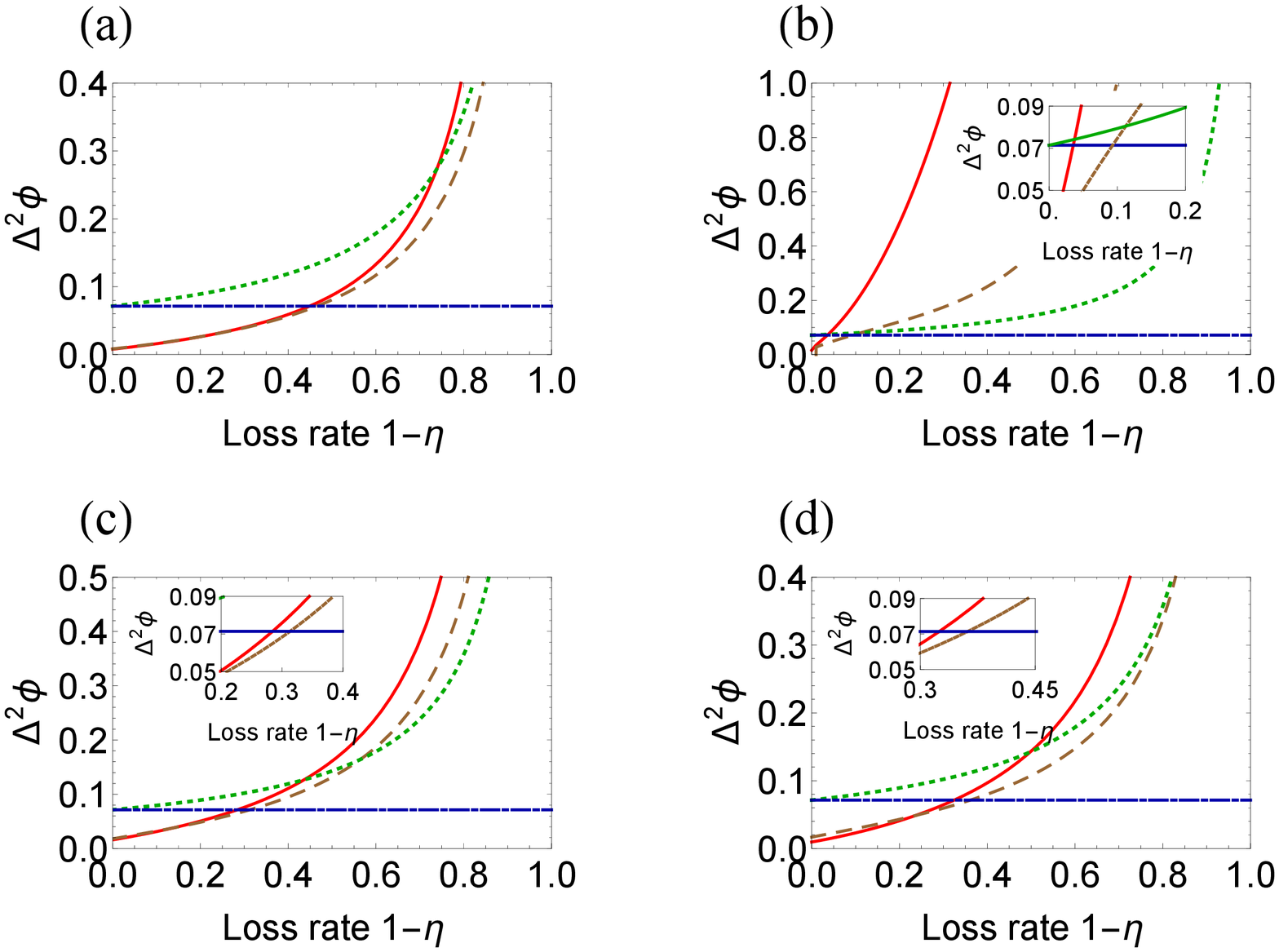}}}
\caption{
Phase sensitivities for CSV (brown dashed curve), TMSV (red solid curve), and coherent (green dotted curve) states, respectively, as a function of loss rate $1-\eta$ 
at $\bar{n}=7$ under photon-loss in one arm via (a) the quantum Cram\'{e}r-Rao bound, (b) parity measurement with PNRD, (c) quadrature measurement with single homodyne detection, and (d) quadrature measurement with double homodyne detection.
The shot-noise limit (blue dot-dashed line) is given by $\Delta^2\phi_{\text{SNL}}=1/2\bar{n}$.
}
\end{figure}

We show phase sensitivities with the mean photon number $\bar{n}=7$, which is implementable with single-mode squeezed vacuum states under the current technology \cite{vahlbruch2016}. 
Figure A1 shows phase sensitivities under symmetric photon loss and Fig. A2 shows those under photon loss on one arm. Both figures exhibit the similar tendency for the case with $\bar{n}=10$, which we presented in the main text.  

\setcounter{equation}{0}
\renewcommand{\theequation}{B\arabic{equation}}
\subsection{Gaussian state}
The MZI dynamics transforms the covariance matrix and the first-order moment of an input Gaussian state as
\begin{align}
\gamma&\xrightarrow{\text{MZI}} \gamma '=M_{\hat{U}_\text{MZI}}\gamma M_{\hat{U}_\text{MZI}}^T, \label{trans1} \\
d&\xrightarrow{\text{MZI}} d'=M_{\hat{U}_\text{MZI}} d, \label{trans2} 
\end{align}
where $M_{\hat{U}_\text{MZI}}=M_{\hat{B}_2}M_{\hat{P}_\phi}M_{\hat{B}_1}$ is the symplectic transformation matrix of the MZI dynamics composed of two 50:50 beam splitters and a phase shifter. In the lossy MZI, the loss channel is inserted between the first beam splitter and the phase shifter as
\begin{align}
\gamma\xrightarrow{\text{MZI}}\gamma'&= M_{\hat{B}_2}M_{\hat{P}_\phi}\mathcal{N}\big(M_{\hat{B}_1}\gamma M^T_{\hat{B}_1}\big)M^T_{\hat{P}_\phi}M^T_{\hat{B}_2}, \label{lossymzi1} \\
d\xrightarrow{\text{MZI}}d'&=M_{\hat{B}_2}M_{\hat{P}_\phi}D_1M_{\hat{B}_1}d, \label{lossymzi2} 
\end{align}
where $\mathcal{N}$ represents the transformation of the covariance matrix by the photon-loss channel.
A detailed analysis of Gaussian states is given in the Refs \cite{ferraro2005, wang2007, marian2012, banchi2015}.

For the symplectic transformation matrix of the MZI dynamics $M_{\hat{U}_\text{MZI}}=M_{\hat{B}_2}M_{\hat{P}_\phi}M_{\hat{B}_1}$, we consider the following formula.
The symplectic matrix that corresponds to the beam splitter is given by
\begin{align}
M_{\hat{B}}(\theta)=
\begin{pmatrix}
\cos\theta & 0 & \sin\theta & 0 \\
0 & \cos\theta & 0 & \sin\theta \\
-\sin\theta & 0 & \cos\theta & 0 \\
0 & -\sin\theta & 0 & \cos\theta
\end{pmatrix}.
\end{align}
The first and second beam splitters correspond to $M_{\hat{B}_1}=M_{\hat{B}}(\pi/4)$ and $M_{\hat{B}_2}=M_{\hat{B}}(-\pi/4)$, respectively. Fictitious beam splitters that describe photon loss can also be described with the symplectic matrix with $\theta=\arccos\sqrt{\eta}$, where $\eta$ is a transmissivity. For the fictitious beam splitters, the first two columns and rows represent the mode that we consider, and the last two columns and rows represent the mode of the environment. Consequently, it can be found that the photon-loss channel with transmissivities $\eta_a$ and $\eta_b$ for modee $a$ and $b$ transforms the covariance matrix and the first-order moment of two-mode Gaussian states as

\begin{align}
\gamma&\xrightarrow{\text{loss}}\mathcal{N}(\gamma)=D_1\gamma D_1+D_2/2, \label{covloss} \\
d&\xrightarrow{\text{loss}}D_1d, \label{fmloss}
\end{align}
where $D_1=\text{diag}(\sqrt{\eta_a},\sqrt{\eta_a},\sqrt{\eta_b},\sqrt{\eta_b})$ and $D_2=\text{diag}(1-\eta_a,1-\eta_a,1-\eta_b,1-\eta_b)$ are diagonal matrices.
The symplectic matrix of the phase shifter operator is given by
\begin{align}
M_{\hat{P}_\phi}=
\begin{pmatrix}
\cos\phi & \sin\phi & 0 & 0 \\
-\sin\phi & \cos\phi & 0 & 0 \\
0 & 0 & 1 & 0 \\
0 & 0 & 0 & 1
\end{pmatrix}.
\end{align}




\subsection{Calculation of Quantum Fisher Information}
\setcounter{equation}{0}
\renewcommand{\theequation}{C\arabic{equation}}
The Bures fidelity of two-mode Gaussian states $\hat{\rho}_1$ and $\hat{\rho}_2$ is given by \cite{marian2012, banchi2015}
\begin{align}
\mathcal{F}(\hat{\rho}_1,\hat{\rho}_2)=\mathcal{F}_0(\gamma_1,\gamma_2)\exp\bigg[-\frac{1}{4}\delta_{d}^T(\gamma_1+\gamma_2)^{-1}\delta_{d}\bigg],
\end{align}
where $\{\gamma_1, d_1\}$ and $\{\gamma_2, d_2\}$ are the covariance matrices and first-order moments of $\rho_1$ and $\rho_2$, respectively. Note that $\delta_d=d_2-d_1$, $\mathcal{F}_0(\gamma_1,\gamma_2)=[\sqrt{\Gamma}+\sqrt{\Lambda}-\sqrt{(\sqrt{\Gamma}+\sqrt{\Lambda})^2-\Delta}]^{-1/2}$, $\Delta=\det(\gamma_1+\gamma_2)$, $\Gamma=2^{4}\det(\Omega \gamma_1 \Omega \gamma_2-\mathbb{1}/4)$, and $\Lambda=2^{4}\det(\gamma_1+i\Omega/2)\det(\gamma_2+i\Omega/2)$. 
Using the covariance matrix and first-order moment with a parameter $\phi$, we calculate the QFI as
\begin{align}
F_Q=\frac{8[1-\mathcal{F}(\hat{\rho}_\phi,\hat{\rho}_{\phi+d\phi})]}{d\phi^2}.
\end{align}

Using the formula (C2), we obtain the QFIs of CSV and TMSV states in the lossy MZI.
For symmetric photon-loss with transmissivities ($\eta_a=\eta_b=\eta$),
\begin{align}
F_Q^{\text{CSV}}&=\frac{\eta\sinh^2{r}[4\eta-1+2\eta^2(3-2\eta)\sinh^2{r}]}{1+2\eta(1-\eta)\sinh^2{r}} \\
&+\frac{2\alpha^2\eta(e^r-\eta\sinh{r})}{e^r-2\eta\sinh{r}}, \nonumber\\
F_Q^{\text{TMSV}}&= \frac{2 \eta^2 \sinh^2{2s}}{1+2\eta(1-\eta)\sinh^2{s}}. \label{tmsv1}
\end{align}
For photon loss only in one arm with transmissivity ($\eta_a=\eta$ and $\eta_b=1$) on which the phase shifter exists,
\begin{align}
F_Q^{\text{CSV}}&=2\eta\bigg(\frac{\sinh^2{r}}{1+\eta}+\frac{\alpha^2\cosh{r}}{\cosh{r}-\eta\sinh{r}} \\
&+\frac{\eta\sinh^2{2r}}{3+\eta^2+(1-\eta^2)\cosh{2r}}\bigg),\nonumber \\
F_Q^{\text{TMSV}}&= \frac{2 \eta^2 \sinh^2{2s}}{1+2\eta(1-\eta)\sinh^2{s}}. \label{tmsv2}
\end{align}
Note that the photon loss that occurs in the empty arm does not change the QFI such that Eqs. (C4) and (C6) are the same.
In Fig. 1, a TMSV state is transformed to a product state of two single-mode squeezed vacuum states after the first 50:50 beam splitter. Then the phase information is encoded only in one of two single-mode squeezed vacuum states. Since the state is a product state even after loss channels, the phase encoded single-mode squeezed vacuum state is not influenced by the other single-mode squeezed vacuum state. Thus, the QFI is calculated by using one of the two single-mode squeezed vacuum states that contains phase information.


\begin{references}

\bibitem{giovannetti2004} V. Giovannetti, S. Lloyd, and L. Maccone, Science \textbf{306}, 1330 (2004).

\bibitem{abbott2016} B. P. Abbott \textit{et al.} (LIGO Scientific and Virgo Collaborations), Phys. Rev. Lett. \textbf{116}, 061102 (2016).

\bibitem{dowling2008} J. P. Dowling, Contemporary Physics, \textbf{49}, 125 (2008).

\bibitem{caves1981} C. M. Caves, Phys. Rev. D \textbf{23}, 1693 (1981).

\bibitem{braunstein1994} S. L. Braunstein and C. M. Caves, Phys. Rev. Lett. \textbf{72}, 3439 (1994).

\bibitem{paris2009} M. G. A. Paris, Int. J. Quantum Inf. \textbf{7}, 125 (2009).

\bibitem{joo2011} J. Joo, W. J. Munro, and T. P. Spiller, Phys. Rev. Lett. \textbf{107}, 083601 (2011).

\bibitem{joo2012} J. Joo, K. Park, H. Jeong, W. J. Munro, K. Nemoto, and T. P. Spiller, \pra~\textbf{86}, 043828 (2012).

\bibitem{lee2015} S.-Y. Lee, C.-W. Lee, H. Nha, and D. Kaszlikowski, J. Opt. Soc. Am. B \textbf{32}, 1186 (2015).

\bibitem{zhang13} Y. R. Zhang, G. R. Jin, J. P. Cao, and W.M. Liu, J. Phys. A \textbf{46}, 035302 (2013).

\bibitem{knott2016} P. A. Knott, T. J. Proctor,  A. J. Hayes, J. P. Cooling, and J. A. Dunningham, \pra~\textbf{93}, 033859 (2016).

\bibitem{lee2016} S.-Y. Lee, C.-W. Lee, J. Lee, and H. Nha, Sci. Rep. \textbf{6}, 30306 (2016).

\bibitem{escher2011} B. M. Escher, R. L. de Matos Filho, and L. Davidovich, Nat. Phys. \textbf{7}, 406 (2011).

\bibitem{demkowicz2015} R. Demkowicz-Dobrza{\'n}ski, M. Jarzyna, and J. Ko{\l}ody{\'n}ski, Prog. Opt. \textbf{60}, 345 (2015).

\bibitem{huver2008} S. D. Huver, C. F. Wildfeuer, and J. P. Dowling, \pra~\textbf{78}, 063828 (2008).

\bibitem{datta2011} A. Datta, L. Zhang, N. Thomas-Peter, U. Dorner, B. J. Smith, and I. A. Walmsley, Phys. Rev. A \textbf{83}, 063836 (2011).

\bibitem{dorner2009} U. Dorner, R. Demkowicz-Dobrza{\'n}ski, B. J. Smith, J. S. Lundeen, W. Wasilewski, K. Banaszek, and I. A. Walmsley, Phys. Rev. Lett. \textbf{102}, 040403 (2009).

\bibitem{demkowicz2009} R. Demkowicz-Dobrza{\'n}ski, U. Dorner, B. J. Smith, J. S. Lundeen, W. Wasilewski, K. Banaszek, and I. A. Walmsley, Phys. Rev. A \textbf{80}, 013825 (2009).

\bibitem{kacprowicz2010} M. Kacprowicz, Demkowicz-Dobrza{\'n}ski, W. Wasilewski, K. Banaszek, and I. A. Walmsley, Nat. Photon. \textbf{4}, 357 (2010).

\bibitem{pezze2008} L. Pezz\'e and A. Smerzi. Phys. Rev. Lett. \textbf{100}, 073601 (2008).

\bibitem{seshadreesan2011} K. P. Seshadreesan, P. M. Anisimov, H. Lee, and J. P. Dowling, New J. Phys \textbf{13}, 083026 (2011).

\bibitem{jarzyna2012} M. Jarzyna and R. Demkowicz-Dobrza{\'n}ski, Phys. Rev. A \textbf{85}, 011801 (2012).

\bibitem{lang2013} M. D. Lang and C. M. Caves, Phys. Rev. Lett. \textbf{111}, 173601 (2013).

\bibitem{DBS2013} R. Demkowicz-Dobrza{\'n}ski, K. Banaszek, and R. Schnabel, \pra~\textbf{88}, 041802(R) (2013).

\bibitem{Gao2016} Y. Gao, \pra \textbf{94}, 023834 (2016).

\bibitem{gard2017} B. T. Gard, C. You, D. K. Mishra, R. Singh, H. Lee, T.R. Corbitt, and J. P. Dowling, EPJ Quantum Technology \textbf{4}, 4 (2017).

\bibitem{LWYJS17} P. Liu, P. Wang, W. Yang, G.R. Jin, and C.P. Sun, \pra~\textbf{95}, 023824 (2017).


\bibitem{steuernagel2004} O. Steuernagel and S. Scheel, J. Opt. B \textbf{6}, S66 (2004).

\bibitem{olivares2009} S. Olivares and M. G. A. Paris, J. Phys. B \textbf{42}, 055506 (2009).

\bibitem{anisimov2010} P. M. Anisimov, G. M. Raterman, A. Chiruvelli, W. N. Plick, S. D. Huver, H. Lee, and J. P. Dowling, Phys. Rev. Lett. \textbf{104}, 103602 (2010).

\bibitem{plick2010} W.N. Plick, P.M. Anisimov, J.P. Dowling, H. Lee, and G.S. Agarwal, New J. Phys. \textbf{12}, 113025 (2010).

\bibitem{zhang2013} X.-X. Zhang, Y.-X Yang, and X.-B. Wang, Phys. Rev. A \textbf{88}, 013838 (2013).

\bibitem{huang2016} Z. Huang, K. R. Motes, P. M. Anisimov, J. P. Dowling, and D. W. Berry,  \pra~\textbf{95}, 053837 (2017).

\bibitem{leonhardt98} U. Leonhardt, \textit{Measuring the Quantum State of Light} (Cambridge University Press, Cambridge, 1998).

\bibitem{ferraro2005} A. Ferraro, S. Olivares, and M. G. A. Paris, \textit{Gaussian States in Quantum Information} (Bibliopolis, Berkeley, 2005).

\bibitem{wang2007} X.-B Wang, T. Hiroshima, A. Tomita, and M. Hayashi, Phys. Rep. \textbf{448}, 1 (2007).

\bibitem{marian2012} P. Marian and T. A. Marian, \pra~\textbf{86}, 022340 (2012).

\bibitem{banchi2015} L. Banchi, S. L. Braunstein, and S. Pirandola, Phys. Rev. Lett. \textbf{115}, 260501 (2015).

\bibitem{sparaciari2015} C. Sparaciari, S. Olivares, and M. G. A. Paris, J. Opt. Soc. Am. B \textbf{32}, 1354 (2015).


\bibitem{barnett2002} S. M. Barnett and P. M. Radmore, \textit{Methods in Theoretical Quantum Optics}, Vol. 15 (Oxford University Press, New York, 2002), Vol. 15.

\bibitem{seshadreesan2013} K. P. Seshadreesan, S. Kim, J. P. Dowling, and H. Lee, Phys. Rev. A \textbf{87}, 043833 (2013).

\bibitem{gardiner1991} C. W. Gardiner and H. Haken, \textit{Quantum Noise} (Springer Berlin, 1991), Vol. 26.

\bibitem{vahlbruch2016} H. Vahlbruch, M. Mehmet, K. Danzmann, and R. Schnabel, Phys. Rev. Lett. \textbf{117}, 110801 (2016).

\bibitem{BW00} D. W. Berry and H. M. Wiseman, \prl~\textbf{85}, 5098 (2000).




\end{references}
\end{document}